# Optical Link Tomography: First Field Trial and 4D Extension

Takeo Sasai, *Member, IEEE*, Giacomo Borraccini, *Member, IEEE*, Yue-Kai Huang, *Member, IEEE*, Hideki Nishizawa, Zehao Wang, *Student Member, IEEE*, Tingjun Chen, *Member, IEEE*, Yoshiaki Sone, Minami Takahashi, Tatsuya Matsumura, Masanori Nakamura, *Member, IEEE*, Etsushi Yamazaki, *Member, IEEE*, Koichi Takasugi, Ting Wang, and Yoshiaki Kisaka

(*Post-Deadline Paper*)

*Abstract*— **Optical link tomography (OLT) is a rapidly evolving field that allows the multi-span, end-to-end visualization of optical power along fiber links in multiple dimensions from network endpoints, solely by processing signals received at coherent receivers. This paper has two objectives: (1) to report the first field trial of OLT, using a commercial transponder under standard DWDM transmission, and (2) to extend its capability to visualize across 4D (distance, time, frequency, and polarization), allowing for locating and measuring multiple QoT degradation causes, including time-varying power anomalies, spectral anomalies, and excessive polarization dependent loss. We also address a critical aspect of OLT, i.e., its need for high fiber launch power, by improving power profile signal-to-noise ratio through averaging across all available dimensions. Consequently, multiple loss anomalies in a field deployed link are observed even at launch power lower than system optimal levels. The applications and use cases of OLT from network commissioning to provisioning and operation for current and near-term network scenarios are also discussed.**

## I. INTRODUCTION

The availability of distributed parameters within systems or objects is highly beneficial for their inspection and design. Tomographic approaches provide non-destructive measurements of internal parameter distributions using externally observed data with reduced downtime, cost, and safety risks and therefore have long been extensively developed and applied across numerous disciplines, as exemplified by X-ray computed or optical coherence tomography for biomedical imaging [1][2] and industrial diagnostics [3] and seismic tomography for geographical exploration [4].

Also in fiber-optic communication systems, which is a pillar of the modern information era, the monitoring and management of distributed parameters such as optical power along fibers is crucial to fully utilize the potential data capacity of a given link with less system margins and to facilitate fault localization and network automation. In particular, optical power is one of the most critical and variable link parameters that affect the total system capacity as it predominantly determines the generalized signal-to-noise ratio (SNR).

To date, be it distributed or non-distributed, numerous efforts have been devoted to establish and enhance network monitoring functions in the field of optical performance monitoring [5][6]. In current systems, a straightforward approach for physical layer monitoring is to rely on dedicated measuring instruments such as optical time domain reflectometer (OTDR). While this approach generally provides accurate and reliable measurement, the availability of such hardware devices for real-time monitoring is often scarce in legacy networks, and distributing these devices at nodes around networks may not be a reasonable option as it introduces considerable time and cost. Another approach is to use coherent DSP-based monitor [7]. This approach is cost-effective as it imposes no need for deploying additional hardware devices on operators and can monitor a variety of performance metrics simply by accessing transceivers, such as chromatic dispersion (CD) [8], differential group delay (DGD), polarization dependent loss (PDL), and state of polarization. These metrics, however, represent cumulative values over the entire link and lack spatial resolution, limiting the ability to localize performance bottlenecks. Our approach is classified to the latter, i.e., coherent DSP-based monitor, but allows for distributed monitoring.

Moreover, given that modern networks are becoming increasingly complex due to the adoption of multi-band transmission, distributed amplification, and open, disaggregated architectures, these conventional monitoring solutions are facing limitations. In systems employing multi-band (beyond C band) or distributed Raman amplification (DRA), the management of spatially and spectrally distributed

Manuscript received XXXX; Date of publication XXXX; Date of current version XXXX. *(Corresponding author: Takeo Sasai.)*

T. Sasai, H. Nishizawa, Y. Sone, M. Takahashi, T. Matsumura, M. Nakamura, E. Yamazaki, K. Takasugi and Y. Kisaka are with the Network Innovation Laboratories and also with the Device Innovation Center, NTT, Inc., Kanagawa 239-0847, Japan (e-mail: takeo.sasai@ntt.com; hideki.nisizawa@ntt.com; yoshiaki.sone@ntt.com; minami.takahashi@ntt.com; tatsuya.matsumura@ntt.com; msnr.nakamura@ntt.com; etsushi.yamazaki@ntt.com; koichi.takasugi@ntt.com; yoshiaki.kisaka@ntt.com.)

G. Borraccini, Y.K. Huang, and T. Wang are with the NEC Laboratories America, Princeton, NJ 08540 USA (e-mail: gborraccini@nec-labs.com; kai@nec-labs.com; ting@nec-labs.com).

Z. Wang and T. Chen are with the Department of Electrical and Computer Engineering, Duke University, Durham NC 27708 USA (e-mail: zehao.w@duke.edu; tingjun.chen@duke.edu).

Color versions of one or more of the figures in this paper are available online at http://ieeexplore.ieee.org.

Digital Object Identifier XXX

optical power is mandatory due to the power transition by the Raman process: otherwise, the expected SNR or capacity gains cannot be achieved [9][10]. Nevertheless, there is no established method employed in commercial networks to directly measure such an affected spatial and spectral power evolution during operation. In open and disaggregated networks [11], where multi-vendor equipment coexists, there is an increased risk of variability in link parameters or inaccessibility to certain components from end users. This is particularly true in multi-domain networks, which can be seen in the context of optical spectrum as a service scenarios [12] and open all-photonic network [13]. In this architecture, multiple optical network domains function as a single entity without OEO conversion at domain edges, providing end users with end-to-end optical connection with high-capacity, low-latency communication while reducing operators' costs and power consumption. However, due to security concerns, operators and end users have extremely limited access to link and node information beyond their administrative boundaries, which prevents the optimal design of end-to-end lightpath and link margins, the identification of performance bottleneck, and the attribution of responsibility for network failures—particularly for soft failures—thus potentially hindering the widespread adoption of such networks.

Therefore, a *tomographic* approach is desirable, which reconstructs internal distributed properties of the entire link solely from network endpoints (i.e., transponder) without accessing node equipment or additional probing light. Recent research activities have put considerable efforts on developing digital longitudinal monitoring or optical link tomography (OLT) since 2019 [14][15][16][17]. The following is a list of key features of this approach:

- Visualizing multi-span, end-to-end optical power along the fiber link by processing received signal samples at coherent receivers.
- Operating on live signals, allowing simultaneous communication and sensing without interrupting network operations.
- Requiring no extra probing lights nor optical reconfiguration of existing networks, imposing minimal or no hurdles for operators to deploy.

Due to these features, OLT provides a scalable and cost-effective telemetry, regardless of intermediate network configurations, vendors, or domains. It is therefore rapidly evolving and has been discussed from various aspects, ranging from theory [18][19][20] to implementation consideration [21][22][23], and network applications [25][26][27].

The OLT utilizes fiber nonlinearity and chromatic dispersion to achieve distance-resolved monitoring of optical power. Fundamentally, tomography involves solving an inverse problem of estimating distributed parameters from boundary conditions (i.e., Tx and Rx signals in our case) of a system. Such inverse problems are often ill-posed and nontrivial to solve. However, in optical fiber transmission, due to the presence of chromatic dispersion and adequately weak but sufficient fiber nonlinearity, this tomographic problem becomes well-posed and can be uniquely solved in certain conditions, which are usually satisfied in a standard coherent transmission in this era.

Previous studies have expanded the capability of this tomographic approach. The concept of "2D link tomography", which acquires both spatial and spectral power profiles by performing 1D (spatial) tomography using multiple channels, was first demonstrated in [28]. This was followed by a proposal of using it for a network-wide perspective [21], and further extended to C+L [29][30] and S+C+L bands [31]. The extension to polarization dimensions has also been demonstrated, enabling the localization of PDL/polarization dependent gain (PDG) [32][33][34][35]. Combining these dimensions with the additional one—time—may unlock monitoring capabilities and use cases that both conventional measuring instruments and OLT cannot provide.

Furthermore, despite considerable progress, most demonstrations of OLT relied on high fiber launch power to induce sufficient nonlinearity and to achieve high power-profile SNR. This poses challenges for practical deployment of OLT in commercial networks, as it may induce excessive nonlinearity on sensing channel itself and adjacent wavelength division multiplexed (WDM) channels.

This work aims to increase and demonstrate the feasibility of OLT and expand its capability to broaden uses cases of OLT for networking applications. Specifically, we

- report the first field demonstration of OLT using a commercial transponder.
- extend the dimension of OLT to 4D: distance, time, frequency, and polarization, allowing for locating and measuring multiple QoT degradation causes, including time-varying power anomalies, spectral anomalies, and excessive PDL.
- demonstrate an SNR enhancement of OLT by averaging over all available dimensions, enabling the tomography to operate at power lower than the system optimal levels.
- discuss applications of OLT across the network lifecycle from commissioning to provisioning and operation stage of networks, with scenarios including multi-band transmission, DRA, and multi-domain networks.

This paper extends our previous work [36] by providing the details on the working principle of OLT, a power-profile SNR enhancement strategy validated through simulation, and potential networking applications. The remainder of this paper is organized as follows. Section II describes the system configuration and the working principle of 4D OLT. Section III presents the SNR enhancement strategy along with numerical validation and performance quantification. Section IV details the field demonstration of OLT. Section V discusses the networking applications of OLT across commissioning, provisioning and operational stages of optical networks at the physical layer. Section VI concludes the paper.

## II. OPTICAL LINK TOMOGRAPHY

### A. System Configuration

As shown in Fig. 1, our OLT can be implemented assuming standard dense WDM (DWDM) coherent transmission systems. It provides non-disruptive distributed measurement of the optical power (more precisely, the averaged nonlinear phase shifts) across multiple dimensions including distance, polarization, frequency, and time. OLT does not require the use of specialized signals or specific modulation formats, enabling the use of conventional transceivers, although the use of specialized signals can lead to improved performance. Moreover, because OLT is realized by processing signals received at transceivers, it can operate transparently across a wide range of system architectures, including open, multi-vendor and closed, single-vendor configurations as well as in both single- and multi-domain environments. While high fiber launch power is desirable for OLT to excite sufficient fiber Kerr nonlinearity and achieve precise measurement, a system-optimal (SNR-maximizing) power level or lower is used in the following field experiments by leveraging the precision enhancement technique discussed in Section III. After reception and demodulation, including CD compensation (CDC), frequency offset compensation (FOC), polarization demultiplexing, and carrier phase recovery (CPR), the signals are split into two paths: one is fed to the OLT function block, while the other is used for reconstructing the transmitted signals. The spatial and polarization tomography are then performed using a synchronized pair of these demodulated and reconstructed signals. This process is repeated for consecutively captured signals to achieve temporal tomography. Additionally, the same procedure is applied across multiple WDM channels to obtain spectral tomography.

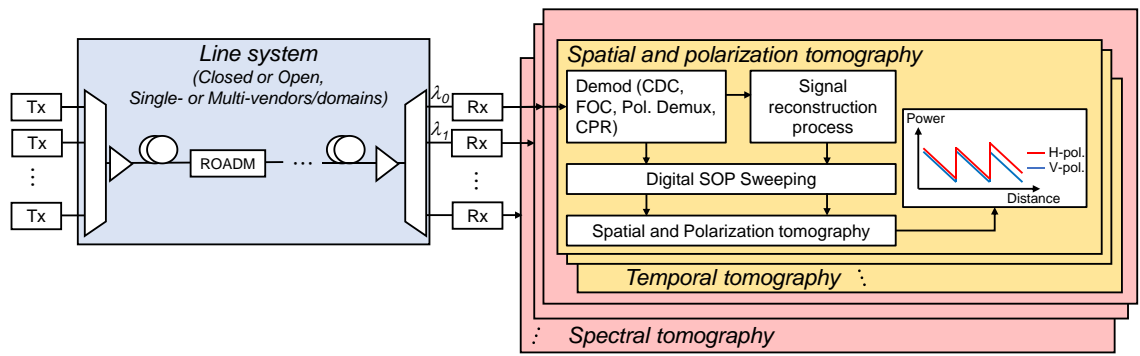

Fig. 1. System configuration of OLT. Spatial and polarization tomography are obtained by processing the received signal of a polarization-multiplexed channel. Temporal tomography is realized by continuously capturing received waveforms and performing spatial (or polarization) tomography over time. Spectral tomography is obtained by applying the above processing to multiple WDM channels of interest.

The essential input for OLT is thus solely the received signal sequences of WDM channels of interest. However, it is often supplemented with prior information of link (or span-wise) distance, which is necessary to obtain the true value of $\gamma'$ and to map it accurately to positions $z$. Although a CD coefficient is also needed for accurate position mapping, it can also be extracted from the received signals and is commonly available from the Rx DSP. Precisely speaking, a full CD map is needed for heterogeneous fiber links; however, this can be computed from prior knowledge of span lengths and the tomography results obtained under the assumption of uniform CD, as demonstrated in [37][38].

### B. Spatial Tomography (Longitudinal Power Monitoring)

Spatial tomography, also referred to as longitudinal power monitoring, reconstructs the optical power profile $P(z)$ along a fiber link from received signals by extracting self-channel nonlinear interference (NLI) (precisely speaking, the position-wise time-averaged nonlinear phase shifts $\gamma'(z) = \gamma(z)P(z)$) imparted to the signals during fiber transmission, where $\gamma(z)$ represents the fiber nonlinear constant at position $z$. In this paper, we refer to this function as "spatial tomography." It represents the most fundamental function, forming the basis for extending tomography to other dimensions. The mechanism enabling the distributed measurement of optical power lies in the interaction between Kerr nonlinearity and CD. As the signal propagates along the fiber, local NLIs are generated at each point in proportion to the local optical power. At the same time, CD continuously alters the signal waveform, such that the NLIs generated at different positions are progressively decorrelated with each other and are thus distinguishable at the receiver. As a result, the total nonlinear distortion observed at the receiver, which is a sum of local effects weighted by the local optical power, can be decomposed due to this decorrelation by solving a linear inverse problem.

A more concrete yet simple formulation is provided below based on [39]. In the first-order regular perturbation model [40][41][42] for nonlinear fiber propagation with a single polarization, the received signal is modeled as $A(L,t) \simeq A_0(L,t) + A_1(L,t)$. Here, $A_0$ represents the linear term obtained by applying only CD to the transmitted signal, and it accounts for the dominant portion of the total received signal $A$. Our interest is $A_1$, which denotes the nonlinear perturbative term, as it carries information on our estimation target $\gamma'(z)$. The modeling of $A_1$ is as follows. During the signal propagation, the additive NLI $-j\gamma'(z)|A_0(z,t)|^2 A_0(z,t)$ is generated locally at each position $z$ along the fiber, depending on the instantaneous (linear) signal waveform $A_0(z,t)$. Note that this local NLI is proportional to our estimation target $\gamma'(z)$. The local NLI propagates toward the receiver as it undergoes residual CD $\widehat{D}_{zL}$ from $z$ to the link end $L$. The received local NLI is thus $\gamma'(z)g(z,t)$, where

$$g(z,t) \equiv -j\widehat{D}_{zL}[|A(z,t)|^2 A(z,t)]. \quad (1)$$

The total NLI $A_1(L,t)$ at the receiver is the accumulation of local NLIs from all positions:

$$A_1(L,t) = \int_0^L \gamma'(z)g(z,t)dz. \quad (2)$$

This is a Fredholm integral equation. In a discretized representation, (2) can be viewed as a linear transformation of our estimation target $\gamma'(z)$ as:

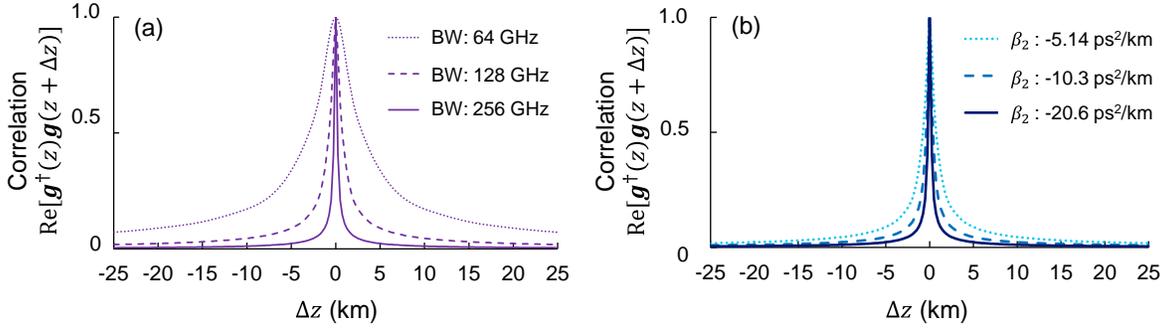

Fig. 2. Spatial correlation function. (a) Dependency on signal bandwidth (BW) and (b) on group velocity dispersion $\beta_2$. In (a), $\beta_2$ is fixed to -20.6 ps$^2$/km while in (b), BW is fixed to 256 GHz. Nyquist-spectral and Gaussian constellation signal is used.

$$A_1(L) = G\gamma'. \quad (3)$$

(2) or (3) means that $g(z, t)$ forms a basis of the total NLI $A_1$. Based on the model (3), the penalized least squares estimator of $\gamma'$ is [39][44]

$$\hat{\gamma}' = (\text{Re}[G^\dagger G] + \lambda R)^{-1} \text{Re}[G^\dagger A_1], \quad (4)$$

where $\lambda$ and $R$ is a regularization parameter and matrix. In our field demonstration, the estimator is extended to dual polarization case, as described in the next subsection.

In general, the inverse problem of this kind estimating distributed parameters solely from system boundaries (i.e., transmitted and received signals) is often ill-posed. In fact, in the case of the spatial tomography, strong correlations arise between local NLIs $g(z, t)$ generated at two adjacent positions when CD is insufficient, making it challenging to distinguish them. Fig. 2 shows the correlation between $g(z, t)$ and $g(z + \Delta z, t)$, called the spatial correlation function or spatial response function [18], which fundamentally governs most of performance metrics of the spatial tomography such as the spatial resolution [18][43], variance, SNR, dynamic range, and minimum detectable loss [20]. For small $\Delta z$, a strong correlation (or non-orthogonality) is observed. This means that it is hard to distinguish these two NLIs at adjacent locations and, consequently, the corresponding two optical power. In contrast, as $\Delta z$ increases, the correlation weakens and the orthogonality is strengthened, implying that two powers can be more easily distinguished. Similarly, the correlation can also be suppressed by increasing signal bandwidth BW (Fig. 2(a)) and group velocity dispersion $\beta_2$ (Fig. 2(b)). These observations come from the fact that the walk-off effect due to CD within $\Delta z$ alters the signal waveform in a fiber, making the excited nonlinearities at adjacent locations decorrelated. This leads to the well-conditioning of the matrix $G$ in (3) and the existence of a stable inverse of it, allowing $\gamma'$ to be reconstructed using (4). For the theoretical aspects and quantification of these fundamentals, readers refer to [20][39].

### C. Polarization Tomography

By extending the spatial tomography for the polarization-multiplexed signal, the polarization-wise power profiles can be estimated, allowing for the distribute measurement of PDL/PDG. The basic principle of polarization tomography is to exploit the nonlinear phase shifts imparted to each polarization component during fiber transmission. Both self-polarization and cross-polarization nonlinearities contribute to these shifts and can thus be used to exploit polarization-wise profiles.

The following provides a formulation analogous to that in the previous section. In this paper, we use a modified version of the linear least-squares estimator proposed in [34] (or Appendix A in [39]), where the model of fiber nonlinear propagation is based on the regular perturbation of the Manakov equation. Based on this model, the linear model (3) can be extended by using a block matrix as follows:

$$\begin{bmatrix} A_{1,x} \\ A_{1,y} \end{bmatrix} \simeq \begin{bmatrix} G_{xx} & G_{xy} \\ G_{yx} & G_{yy} \end{bmatrix} \begin{bmatrix} \gamma'_x \\ \gamma'_y \end{bmatrix}, \quad (5)$$

where $A_{1,i} = A_i(L) - A_{0,i}(L)$ ($i \in \{x, y\}$) is a nonlinear perturbation term in $i$-axis obtained by subtracting the linear term $A_{0,i}(L)$ from received signals $A_i(L)$, $\gamma'_i = \frac{8}{9}\gamma P_i$, $P_i$ is the power of each polarization. $G_{ii}$ and $G_{ij}$ ($i \neq j, j \in \{x, y\}$) are matrices accounting for self- and cross-polarization nonlinearity, whose columns are expressed as

$$\{G_{ii}\}_m = -j\Delta z \, D_{z_m L}\left[\left|A_{i,0}(z_m)\right|^2 A_{i,0}(z_m)\right],$$
$$\{G_{ij}\}_m = -j\Delta z \, D_{z_m L}\left[\left|A_{j,0}(z_m)\right|^2 A_{i,0}(z_m)\right]. \quad (6)$$

Our estimator is modified from the original one [34] in that the penalized least squares with the Tikhonov regularization is introduced to stabilize the solution, which is expressed in the same way as (4):

$$\begin{bmatrix} \hat{\gamma}'_x \\ \hat{\gamma}'_y \end{bmatrix} \simeq \text{Re}\left[\begin{bmatrix} G_{xx} & G_{xy} \\ G_{yx} & G_{yy} \end{bmatrix}^\dagger \begin{bmatrix} G_{xx} & G_{xy} \\ G_{yx} & G_{yy} \end{bmatrix} + \lambda R\right]^{-1} \text{Re}\left[\begin{bmatrix} G_{xx} & G_{xy} \\ G_{yx} & G_{yy} \end{bmatrix}^\dagger \begin{bmatrix} A_{1,x} \\ A_{1,y} \end{bmatrix}\right]. \quad (7)$$

Note that, in this case, optical powers of each polarization are estimated as $\begin{bmatrix} \hat{P}'_x \\ \hat{P}'_y \end{bmatrix} = \frac{9}{8\gamma} \begin{bmatrix} \hat{\gamma}'_x \\ \hat{\gamma}'_y \end{bmatrix}$. Compared to other similar approaches [32][35], this estimator extract information of optical power by exploiting not only self-polarization nonlinearity but also cross-polarization nonlinearity, yielding a power-profile SNR gain.

However, the estimator (7) can only estimate longitudinal power profiles of one specific orthogonal basis $x$ and $y$. In general, the power loss/gain in each polarization due to PDL/PDG depends on the incident state of polarization (SOP) entering a PDL/PDG medium. Consequently, merely estimating the polarization-wise power profiles with (7) may not reveal the presence of PDL/PDG. For instance, in an extreme case where the incident SOP is such that orthogonal polarizations undergo identical losses at a PDL/PDG medium, the resulting polarization-wise power profiles may appear to have a mere lumped loss, rather than a PDL/PDG. To avoid this issue, the basis transformation of received and reference signals is conducted just before the tomography is performed in the Rx DSP chain (see Fig. 1). By doing so, the power profiles of different bases, (e.g., $\frac{1}{\sqrt{2}}(x+y)$ and $\frac{1}{\sqrt{2}}(x-y)$ ) can be estimated. By searching for the basis that maximizes the power difference between two polarization components (i.e., the worst case), the principal axis of PDL /PDG can be identified, allowing for estimating locations and values of PDL/PDGs. Note that, in this scheme, all process is conducted in the Rx DSP and the intentional SOP change in optical domain or at the Tx DSP is unnecessary. The detailed description of this strategy can be found in [32][34]. It has also been demonstrated [34][35] that even when multiple PDLs are present in a link, they can be located simultaneously in a single sweep of bases.

The presence of excess DGDs affects the accuracy of polarization tomography as a timing mismatch between polarizations implies a deviation from the model (5). In [45], distributed measurement of DGDs using a coherent receiver was demonstrated by extending the algorithm of spatial tomography. By incorporating the distributed DGDs estimated by this method into the model, the degradation of polarization tomography due to DGD may be relaxed.

### D. Spectral Tomography

As the spatial and polarization tomography estimates power profiles from the self-channel nonlinear interference, the estimated power is channel-specific. By performing them for all WDM channels available and observing them in the frequency direction, power spectra at arbitrary positions, namely spectral tomography, can be reconstructed without optical channel monitors or spectrum analyzers at intermediate nodes [28][29]. The spectral tomography visualizes how spectral tilts, dips, or other spectral anomalies evolve along the link and thus provides more detailed information on system performance, which has not been possible even with node-based dedicated instruments. This capability is particularly beneficial in systems involving Raman scattering, such as those employing ultra-wideband (UWB) transmission and DRA, where system design can be highly complex due to strong spatial- and wavelength-dependency. A more detailed discussion of this point is provided in Section V.

The frequency resolution of spectral tomography is essentially determined by the channel grid (i.e., tens to hundreds GHz). One approach to enhance the frequency resolution is to split the received signals into several spectra and performing tomography on each split signal. However, this resolution enhancement may come at the cost of reduced accuracy or limited spatial resolution due to reduced power and bandwidth in each split signal. In other words, there should be the performance tradeoff between frequency resolution and estimation accuracy or spatial resolution. Another approach is to embed FIR filters into the channel model, as demonstrated in [16], where gradient-based optimization of the split-step method is used to estimate the in-channel frequency responses at a resolution of a few GHz. This approach enables the identification and localization of abnormal optical filters in a link (e.g., wavelength-selective switches with spectral narrowing or center frequency detuning).

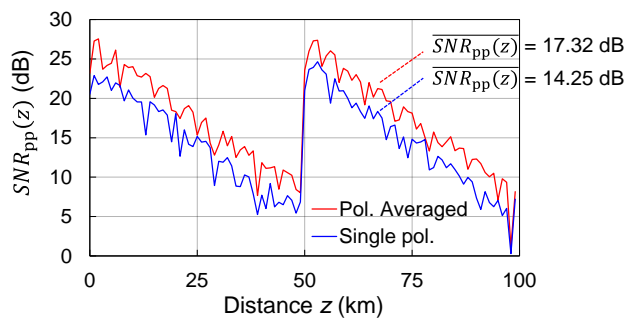

Fig. 3. Power-profile SNR for single-pol. and pol.-averaged estimations. A total of 272 power profiles were used to calculate the statistical $SNR_{pp}(z)$.

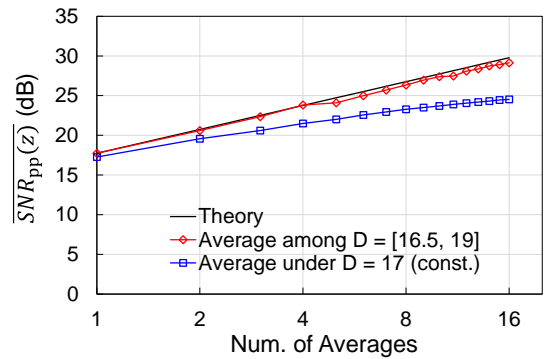

Fig. 4. Averaged power-profile SNR under constant and various CD. When averaging across different D values (blue square), 16 equally spaced points in the range [16.5, 19] were used.

It should be noted that in systems employing semiconductor optical amplifiers (SOAs), the estimated power profiles exhibit an undesired offset due to the SOA nonlinearity, which resembles fiber nonlinear phase rotation, as reported in [46]. In such cases, the simple spatial and spectral tomography may lead to misinterpretation of amplifier behavior. The authors, however, simultaneously proposed incorporating an SOA nonlinearity model into equation (3) to mitigate this offset.

### E. Temporal Tomography

The spatial tomography is performed by capturing and processing received waveforms. By consecutively repeating this process, temporal tomography, which resolves power information over time, is possible. The temporal tomography allows for real-time monitoring of dynamic power variations at arbitrary position, providing localization of, for instance, bending loss caused by mishandling of fibers, eavesdropping, or disaster and gain variation of amplifiers due to channel add/drop or pump degradation. If the power variation is sufficiently slower than the temporal sampling rate, then the system can raise the alarm based on the temporal tomography results, allowing for proactive maintenance before fatal network faults occur. In this way, temporal tomography enhances network resilience against network faults, disruptions, and malicious threats by quickly detecting and localizing power anomalies. Application side of the temporal tomography will also be discussed in Section V.C.

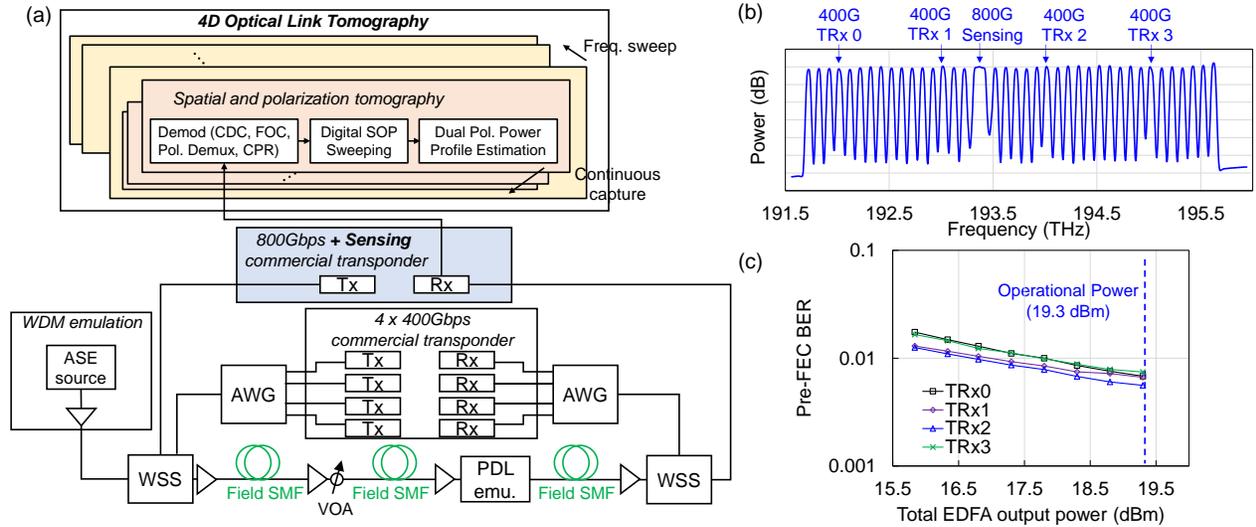

Fig. 5. (a) Field experimental setup. (b) Transmitted spectra. Four 400-Gbps real-time signal are simultaneously transmitted with an 800-Gbps sensing signal. (c) Pre-FEC BER of 4 × 400-Gbps real-time transceivers as a function of total EDFA output power (3-span average). The operational power is fixed at 19.3 dBm, which is below or equal to the level that minimizes the pre-FEC BER.

In our demonstration, we capture waveforms from a transponder every 0.55 sec. This does not always imply that the temporal resolution is 0.55 sec. Power profiles are typically averaged over time to enhance the SNR, and therefore resolution depends on averaging method and size. To improve temporal resolution, the averaging size should be reduced, which in turn limits the SNR improvement, implying that there is a trade-off between temporal resolution and precision. In the following field demonstration, a moving average over three time samples are used, implying resolution is greater than 0.55 sec.

### III. SNR Improvement through Averaging over 4D

The primary challenge of OLT lies in the fact that its estimation principle relies on Kerr nonlinearity, and therefore requires a high fiber launch power for precise estimation. The excessive launch power generates excessive NLI not only on the sensing channel itself but also on adjacent WDM channels, making it prohibitive for use in commercial networks. To enhance the practicality of OLT in real-world networks, achieving a high power profile SNR even at low optical power levels is essential. In this section, we present the improvement of power profile SNR achieved through averaging over 4D in simulation. The experimental demonstration is reported in Section IV.

In the simulation, a dual-polarization (DP)128-GBd PCS-64QAM signal shaped by the Nyquist filter with a roll-off factor 0.1 is used. The symbol sequences are generated using the Mersenne twister with a fixed seed. The tested link is 50 km × 2 span, emulated by the split-step method for the Manakov model, with $\alpha$ = 0.2 dB/km, $D$ = 17 ps/nm/km, $\gamma$ = 1.3 W$^{-1}$km$^{-1}$, an oversampling rate (OS) of 8, and $\Delta z$ = 200 m. The launch power is set to 3.0 dBm/ch. The amplifiers are emulated with gains compensating for the span loss (10 dB) and a noise figure of 5.0 dB. Upon reception, signals are downsampled to 2 OS, followed by CD compensation and a matched filter. The polarization tomography is then performed using (7) with $\Delta z$ = 1 km and $\lambda$ = 0. To assess the improvement of SNR, we used the following definition of power profile SNR, defined in [20]:

$$\mathrm{SNR}_{\mathrm{pp}}(z_m) = \frac{{\gamma'_m}^2}{\mathrm{Var}[\widehat{\gamma'_m}]} \quad (8)$$

It is noteworthy that by adopting this definition, various performance predictions and designs for the tomography become possible, including the specification of the detectable loss limit and dynamic range, and the requirement for critical parameters such as optical launch power and sample size [20].

Theoretically, the power profile SNR (8) is proportional to the sample size used, according to Eq. (14) in [20]. The same effect is expected when averaging over calculated power profiles, assuming that the signal and noise in power profiles are independent among the averaged ensembles. However, if repetitive signals are used for the tomography, as is the case in this paper, the SNR improvement is limited and does not follow the proportionality to the averaging size, due to the signal pattern effect inherent in Kerr nonlinearity [47].

Fig. 3 shows the power profile SNR for the single polarization case and the polarization-averaged case. To calculate the statistical SNR based on (8), a total of 272 power profiles are used. The SNRs averaged across positions are also shown, demonstrating the SNR improvement of approximately 3 dB by averaging over the polarization dimension. This 3-dB improvement is reasonable given that the signal and noise in both polarizations are independent.

The SNR improvement through averaging over time, however, does not increase proportionally with the averaging size as long as the repetitive signals are used. The blue line in

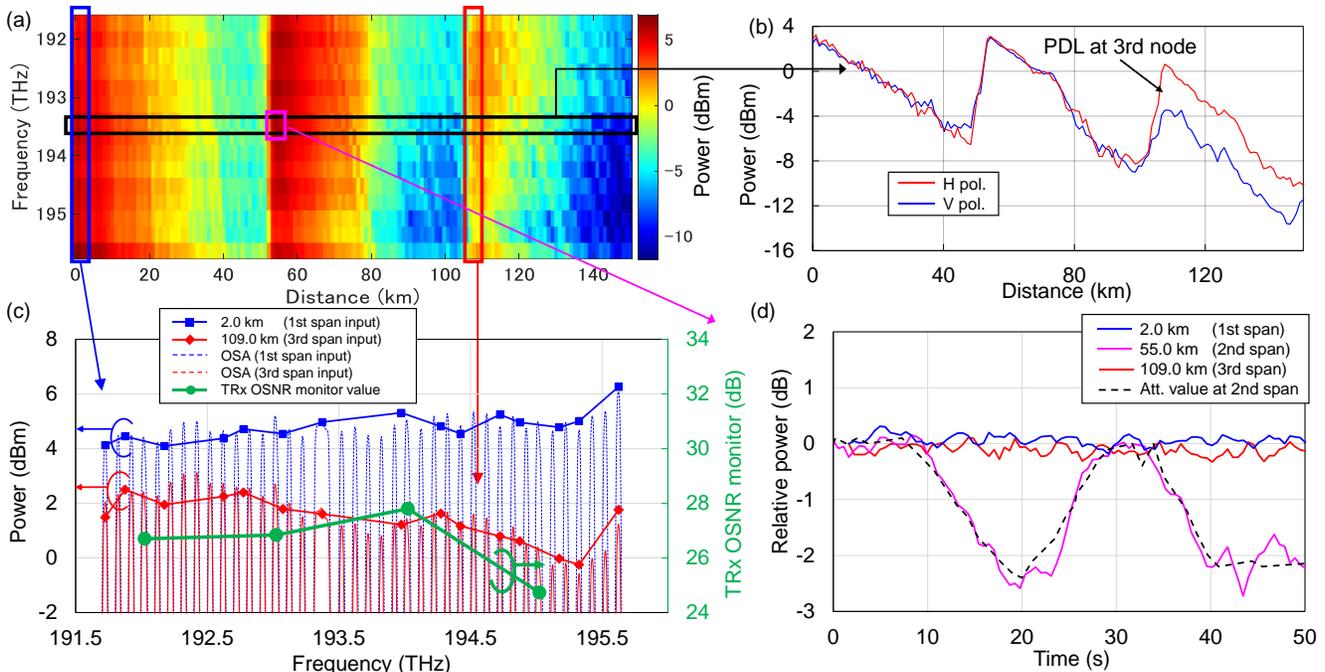

Fig. 6. 4D link tomography results. (a) Spectral and spatial tomography: power profiles obtained at frequencies in C-band. (b) Polarization and spatial tomography: power profiles of each polarization, reflecting the presence of PDL emulator inserted at the beginning of the third span. (c) Spectral tomography at first and third span input. Reference trances obtained by OSA are also shown. OSNR monitored by 4 × 400Gbps TRx are also shown. (d) Temporal tomography: power variations at the span inputs, reflecting the dynamic attenuation applied by the VOA inserted at the beginning of the second span.

Fig. 4 shows the power profile SNR as a function of averaging size over different seeds used for noise, emulating the time averaging. In this case, we fixed CD at $D = 17$ ps/nm/km. An SNR improvement is indeed observed, although it does not increase as rapidly as the theoretical lines, which grows linearly with the averaging size. This discrepancy is due to the signal pattern effect arising from the use of identical repeated signals in generating the power profiles. However, by averaging power profiles over wavelengths, this limitation is relaxed even when the same signal patterns are shared among wavelengths. This is observed in the red line in Fig. 4, where power profile averaging was performed over different CD values to emulate averaging across wavelengths. We used 16 equally spaced CD values, ranging from 16.5 to 19, to generate the 16 power profiles. The SNR improvement aligns with the theoretical expectations. This result can be attributed to the fact that, under different chromatic dispersions, the signal pattern changes at the same location in the optical fiber. Consequently, any pattern-induced effects are effectively averaged out during the averaging process.

## IV. FIELD EXPERIMENT

### A. Experimental Setup

Fig. 5 shows the experimental setup. The test link consists of three spans of field-deployed fibers with 53.4 km, 54.8 km, and 54.8 km, installed around Duke University, Durham, NC. An 800-Gbps signal from a commercial transponder employing a 130-GBd-class DP PCS-64QAM is used for a sensing channel. Additionally, four 400-Gbps real-time data channels (64-GBd-class DP 16QAM signals) are launched simultaneously at frequencies of 192.025, 193.025, 194.025, and 195.025 THz. These real-time data channels are first multiplexed by an arrayed-waveguide grating (AWG) with fixed 100 GHz spacing and then again multiplexed with the sensing channel and emulated DWDM channels by a wavelength selective switch. This two-step multiplexing is due to the broad bandwidth of the sensing channel (130-GBd class) exceeding the fixed channel spacing of the AWG. A fully-loaded 39-channel DWDM system, spaced at 100 GHz across the C-band, is emulated by shaping amplified spontaneous emission (ASE) noise with the WSS. Fig. 5(b) shows the optical spectrum after the first booster erbium-doped fiber amplifier (EDFA). Unlike most previous demonstrations that employed a higher fiber launch power for the sensing channel than the other WDM channels, we set the launch power to the same level as its neighboring channels. The field-deployed fibers had an average CD value of 17.55 ps/nm/km. To demonstrate the localization of a time-varying loss and excessive PDL, a variable optical attenuator (VOA) and an around 3-dB PDL emulation were placed at the beginning of the second and third span, respectively.

For the 400-Gbps channels, pre- and post-FEC bit error rates (BERs) were obtained from the real-time transponder. Fig. 5(c) presents the pre-FEC BER of the four transceivers plotted against the total EDFA output power. The EDFAs were operated in a constant output power mode, and the same output power was applied to all three spans. Based on these results, we selected a launch power of 19.3 dBm, where the pre-FEC BERs still have not reached their optimal peak, ensuring the system remains in a linear regime. This low power operation was due to the limited maximum power of our EDFAs. We observed that the corresponding post-FEC BERs were error-free.

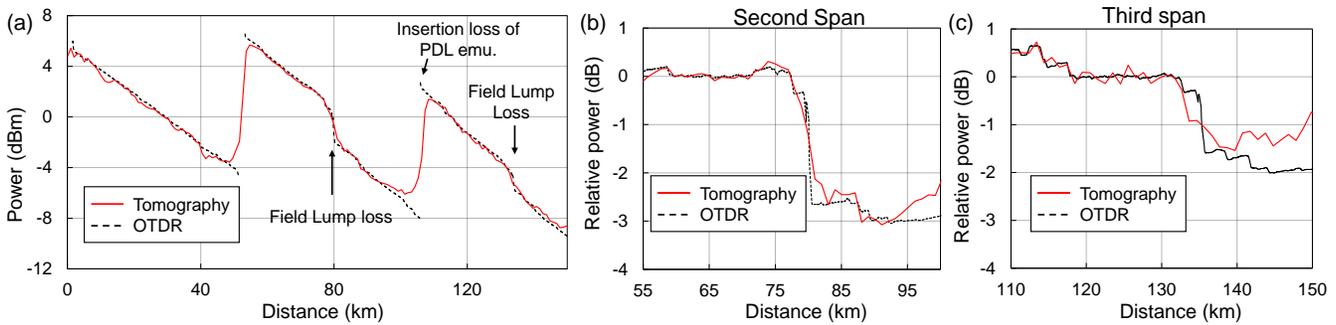

Fig. 7. (a) Spatial tomography results averaged over time, polarization, and frequency dimensions. (b)(c) Closeup of power profiles around field lumped losses in second and third spans, respectively. Fiber loss slopes are subtracted.

For the sensing signal, waveforms are captured from the 800-Gbps transponder at 0.55-second intervals. After reception, the demodulation process including CDC, FOC, polarization demultiplexing, and CPR were conducted. To obtain 4D link tomography results, the polarization tomography was first performed using (7). The temporal tomography was achieved by repeating the polarization tomography for continuously acquired waveforms, with a moving average over three-time samples. The spectral tomography is enabled by varying the center frequency of the sensing channels and conducting the polarization tomography at each frequency. The SOP of both the demodulated and reference signals are then swept to identify the principal axis of the targeted PDL [32][34], as shown in Fig. 1.

## B. 4D Optical Link Tomography Results

Fig. 6 shows the 4D link tomography results. Fig. 6(b) shows the spatial and polarization tomography results based on (7). Throughout the first and second spans, the two orthogonal polarization components maintain identical power levels. However, at the beginning of the third span, where the PDL emulator is placed, noticeable discrepancies were observed in the polarization-wise powers, demonstrating the localization of the PDL. From the spatial and spectral tomography depicted in Fig. 6(a), the locations of the EDFAs are clearly identified. It is noticeable that a few diagonal traces appear (for instance, at around 40 km), which can be attributed to spectral tilt induced by the weak Raman effect within the C-band. Fig. 6(c) shows spectral tomography data collected from the inputs of the first and third spans. The WDM spectra measured by an optical spectrum analyzer are also shown for comparison. The tomography and the optical spectrum analyzers(OSA) references show agreement, with a maximum deviation of 0.62 dB and an RMS error of 0.29 dB. At the first-span input (blue), the spectral tomography result exhibits a tilt trending downward toward the lower-frequency side, resulting in a 2-dB power difference at the spectral edges. In contrast, the tomography at the third-span input (red) shows a tilt emphasized in the opposite direction, which is due to the accumulated gain tilt at EDFAs and by the Raman effect. Furthermore, a notable dip appears in the spectrum around 195.325 THz, which coincides with the optical SNR (OSNR) measurements (green) performed by four 400-Gbps transceivers, with TRx 3 at 195.025 THz exhibiting more OSNR degradation. Fig. 6(d) presents the temporal tomography at 2.0 km, 55.0 km, and 109.0 km, with the sensing signal set to 193.4 THz. The power fluctuation observed at 55.0 km (purple trace) aligns with the monitored output of the VOA placed at the beginning of the second span. In contrast, the constant optical power at 2.0 km and 109.0 km is observed, which is attributed to the constant output power operation mode of the EDFAs.

In this way, the OLT provides link information only by using commercial transponders, which even dedicated equipment cannot offer during operations, allowing for pinpointing the source of QoT degradation observed at the conventional transceivers.

## C. Refined Spatial Tomography

Finally, we present the spatial tomography, enhanced by averaging over time (100 waveforms), polarization (two states), and frequency (15 frequencies), as shown in Fig. 7(a). A non-uniform power observed at the third-span input is attributed to the insertion loss of the PDL emulation. This also implies that in the third span, the tomography is operated with a power lower than the system optimal level. The tomography results show agreement with the results of OTDR reference lines and track power changes at loss anomaly in the second and third spans. Fig. 7(b) and (c) are the close-up of these locations with fiber loss slopes subtracted from the results in Fig. 7(a), serving as anomaly indicators. These maps reveal significant lumped losses in the field fibers, as well as indications of smaller events distributed along the spans. Despite the more severe conditions under low power operation in the third span (Fig. 7(c)), the tomography clearly shows the presence of a 1.5 dB lumped loss.

## V. APPLICATIONS

OLT has manifold applications at every stage of network deployment, from commissioning to provisioning and operation. The following discusses the applications enabled by each tomography and the combinations thereof. By leveraging these applications, one can gain benefits beyond simple cost savings (e.g., avoiding the use of dedicated measuring instruments), achieving functionalities that such instruments alone cannot provide.

### A. Commissioning Stage

During the commissioning stage of line systems, OLT enables end-to-end testing of optical power, fiber loss slope, anomaly losses, PDL/PDG, amplification gain, and spectra based on the actual optical power experienced by the channels available, which allows for the final specifications and verification of the established line systems. If the tomography reveals performance bottlenecks or suboptimal components in either dimension as demonstrated in Section IV, the system can be fixed or further optimized by adjusting parameters such as amplification gain or per-channel launch power. Again, note that specialized signals are not necessary for OLT, and any modulation format can be used, implying the commercial transponders intended for operational use can serve as OLT, as demonstrated in our field experiments.

The spatial and spectral tomography is highly synergistic with power or spectral optimization. Power optimization and WDM-spectrum shaping should generally be performed to maximize the overall SNR or throughput. This maximization can commonly be achieved through approaches such as simple power sweeps or model-based approaches [50][51][52], both of which are highly compatible with the tomography. Specifically, power sweeps can be carried out while observing the visualized tomography results, allowing operators to perform the optimization step-by-step as confirming how the actual end-to-end signal power changes during the optimization. In model-based approaches such as those using the Gaussian noise (GN) model [53][54], one first estimates the overall SNR based on the spatial- and spectral tomography results. By parameterizing the launch power or spectrum and solving an optimization problem of maximizing the SNR or total capacity (though may not be a convex optimization in some situations, e.g., see [55]), the optimal working point and spectral shaping can be determined [51]. This approach is particularly synergistic as model-based methods often suffers from unavailability of physical link parameters, and the OLT helps to address such technical gaps. Furthermore, one can further leverage the fact that the tomography inherently estimates the NLI in its estimation process, as demonstrated in [48][49]. Indeed, the tomography results $\gamma'$ can be re-substituted into (3) to recover waveform-level NLI $A_1$, whose statistical operation can provide the SNR for NLI. The optimal launch power can be determined by incorporating this SNR NLI and OSNR into, for instance, so-called 3-dB rule [56]. Other examples of the combination of the tomographic approach and the line system optimization can be seen in [27][57]. However, the above discussion requires knowledge or assumptions about the noise figure (NF) of each amplifier, as well as the connector losses before and after each amplification, which demands a spatial resolution finer than what current tomography provides. Pioneering works for estimating these physical parameters have been proposed in [58][59] for noise localization and in [27] for connector loss estimation.

The above discussion is particularly applicable to systems involving Raman process, such as ones employing UWB transmission and DRA [55][9]. In these systems, controlling the spatial and spectral power profiles is crucial since Raman gain exhibits a wavelength dependence both on the pump's own frequency and the frequency difference between the channel and the pump, thus complicating system design and management. While it is theoretically possible to compute the longitudinal power evolution from launched power spectra, doing so requires detailed physical parameters—such as wavelength-dependent fiber loss, connector losses, wavelength-dependent Raman gain spectra, launched spectrum, pump power, and pump spectrum and wavelength—which are rarely fully available in real-world operation. The tomography, by contrast, directly visualizes the optical power experienced by the signal, regardless of the availability of these physical parameters. Demonstrations of such visualization in a Raman-involved system using the tomography can be seen in [16][28][31][48].

### B. Provisioning Stage

Like the commissioning process, the lightpath-provisioning stage can utilize the visualized optical power-level diagram to estimate the QoT of a given line system, thereby enabling optimal routing and transmission mode selection for a given network and link with less margin. In this process, the so-called design margin discussed in [60] and [61], which arises from the uncertainty or inaccuracy of link parameters input to QoT estimation tools [62], can be minimized as the tomography effectively supplement the lack of such information.

The availability of distributed PDL/PDG information can offer substantial benefits for margin reduction. Currently, because there is no established method to measure the (worst-case) PDL/PDG across the entire link, a relatively large end-to-end PDL of around 4 to 6 dB and its associated penalty is assumed in, for instance, the OpenROADM [63]. If distributed PDLs can be obtained through polarization tomography, these predefined penalty values can be set on a per-link basis, thereby minimizing the margins associated with the uncertainty of PDL/PDGs. However, the iterative process of SOP sweeping and polarization tomography required to estimate the true value of PDL imposes non-negligible computational burden, which should be taken into account when designing computationally efficient implementations in future work.

If per-lightpath power management is allowed in the presence of other lightpaths, the same discussion as in the previous subsection can apply in the provisioning stage. While this operation is often considered tricky as it may affect responses of amplifiers, induce excess nonlinearity, and thus influence the QoT of other lightpaths, the spatial and spectral tomography results can be further leveraged to confirm in advance that such an operation do not significantly impact on other channels by inputting tomography results into a system design tool and estimating the resulting QoT.

From a network perspective, tomography is particularly advantageous in disaggregated or multi-domain networks [12][13]. In such an environment, where multiple network domains coexist, it is extremely difficult to access network nodes beyond operator's administrative domain and obtain their link parameters. As a result, QoT estimation, identification of performance bottlenecks, and determining responsibility in the event of anomalies pose significant challenges. By employing tomography, however, an end-to-end visualization is available, allowing operators to access link parameters outside their administrative domain, just by accessing network endpoints, i.e., transceivers.

*C. Operation and Maintenance Stage*

During the system operation, the temporal tomography and its combination with other tomographies serve as a root cause analysis tool or proactive maintenance tool for time-varying soft failures and should be detected before service level agreement violation or possible hard failures. Time-varying failures include losses caused by mishandling or aging of fibers, disaster, eavesdropping [24], and gain variation of amplifiers due to channel add/drop or pump degradation. By continuously monitoring the longitudinal optical power level and resulting QoT, one can directly identify the location and amount of power change. By doing so, the system margins [60][61], which are associated with these time-varying impairments or network conditions, can be reduced. From initial and current power levels and QoT, one can calculate the consumed and residual margins since the beginning of network operation, allowing operators to take such measures as proactively rerouting the path and simultaneously investigating and resolving the underlying cause.

While this method is highly effective for slowly varying soft failures spanning for minutes or hours, addressing faster variations—such as instantaneous fiber cuts and SOP changes—pose challenges. Although locations of the root causes can be identified once waveforms are collected at the time of event, achieving proactive countermeasures will require accelerating the entire workflow, for instance, from computation to detection and rerouting.

## VI. Conclusion

In this paper, we presented the optical link tomography, which enables comprehensive visualization of spatiotemporal, spectral, and polarization-dependent power profiles along the entire link from network endpoints (i.e., transponders). This functionality can be incorporated into transponders or a controller with minimal modifications by capturing received waveforms and processing it internally or externally. Using a commercial transponder, we reported the first field trial of the OLT in a standard full C-band transmission, successfully identifying various QoT degradation factors, including time-varying losses, spectral tilts and dips, and PDL in a distance-resolved manner. We also demonstrated that averaging over all accessible dimensions improves the SNR of the measured power profile, allowing the observation of multiple lumped losses in field fibers even under low nonlinearity conditions, where the launch power is below the system's optimal operating level. Extended simulations quantify these findings, demonstrating that averaging over polarization yields a 3 dB gain in the power profile SNR. With averaging over time with repetitive signals, the SNR gain eventually saturates. By contrast, averaging across different wavelengths provides the SNR gain in line with the theoretical expectations, owing to the waveform variations induced by chromatic dispersion in fibers.

These demonstrations highlight the feasibility of OLT in practical networks and its broad functionality. Using this tomographic monitoring approach that is independent of specific network or vendor and domain configurations, operators can efficiently understand the current network status with minimal cost by only accessing transponders, and thus facilitate networking from commissioning to provisioning, operation, and maintenance.

ACKNOWLEDGEMENT

This work was supported in part by NSF grants CNS-2211944 and CNS-2330333. We thank the Duke Office of Information Technology (OIT) team and the members of NTT DevIces America Inc. for their support.

REFERENCES

[1] G. N. Hounsfield, "Computed medical imaging." Science 210(4465), 22-28, 1980.
[2] D. Huang et al. "Optical coherence tomography," Science 254(5035), 1178-1181, 1991.
[3] P.J. Withers, C. Bouman, S. Carmignato et al. "X-ray computed tomography," Nat. Rev. Methods Primers 1(18), 2021.
[4] K. Aki, and W. H. K. Lee. "Determination of three‐dimensional velocity anomalies under a seismic array using first P arrival times from local earthquakes: 1. A homogeneous initial model." *Journal of Geophysical research*, 81(23), 4381-4399, 1976.
[5] D. C. Kilper et al., "Optical performance monitoring," J. Lightw. Technol., vol. 22, no. 1, pp. 294-304, Jan. 2004.
[6] Z. Dong, F. N. Khan, Q. Sui, K. Zhong, C. Lu, and A. P. T. Lau, "Optical performance monitoring: a review of current and future technologies," *J. Light. Technol.*, vol. 34, no. 2, pp. 525–543, Jan. 2016.
[7] F. N. Hauske, M. Kuschnerov, B. Spinnler and B. Lankl, "Optical performance monitoring in digital coherent receivers," *J. Lightw. Technol.*, vol. 27, no. 16, pp. 3623-3631, Aug, 2009.
[8] R. A. Soriano, F. N. Hauske, N. G. Gonzalez, Z. Zhang, Y. Ye and I. T. Monroy, "Chromatic dispersion estimation in digital coherent receivers," *J. Lightw. Technol.*, vol. 29, no. 11, pp. 1627-1637, Jun. 2011.
[9] F. Hamaoka et al., "Ultra-wideband WDM transmission in S-, C-, and L-bands using signal power optimization scheme," *J. Lightw. Technol.*, vol. 37, no. 8, pp. 1764-1771, April, 2019.
[10] I. Roberts, J. M. Kahn, J. Harley and D. W. Boertjes, "Channel power optimization of WDM systems following Gaussian noise nonlinearity model in presence of stimulated Raman scattering," *J. Lightw. Technol.*, vol. 35, no. 23, pp. 5237-5249, Dec. 2017.
[11] M. Newland et al., "Open optical communication systems at a hyperscale operator," *J. Opt. Commun. Netw.*, 12, C50-C57, 2020.
[12] K. Kaeval, K. Grobe, and J.-P. Elbers, "Operation of optical spectrum as a service in disaggregated and multi-operator environments," *J. Opt. Commun. Netw.* vol. 17, no. 1, A46-A58, 2025.
[13] Innovative Optical and Wireless Network (IOWN) Global Forum, "Open all-photonic network functional architecture," 2023. https://iowngf.org/wp-content/uploads/formidable/21/IOWN-GF-RD-Open_APN_Functional_Architecture-2.0.pdf
[14] T. Tanimura et al., "Experimental demonstration of a coherent receiver that visualizes longitudinal signal power…," in *Proc. Eur. Conf. Opt. Commun.*, Ireland, Dublin, Sep. 2019, Art. no. PD.3.4.
[15] T. Sasai *et al.*, "Simultaneous detection of anomaly points and fiber types in multi-span transmission links only by receiver-side digital signal processing," in *Proc. Opt. Fiber Commun. Conf. Expo.*, San Diego, CA, USA, Mar. 2020, Art. no. Th1F.1.
[16] T. Sasai, M. Nakamura, E. Yamazaki, S. Yamamoto, H. Nishizawa, and Y. Kisaka, "Digital longitudinal monitoring of optical fiber communication link," *J. Lightw. Technol.*, vol. 40, no. 8, pp. 2390–2408, Jan. 2022.
[17] T. Tanimura, S. Yoshida, K. Tajima, S. Oda, and T. Hoshida, "Fiber-longitudinal anomaly position identification over multi-span transmission